\documentclass[12pt,floats]{article}
\usepackage{mathrsfs}

\usepackage{graphicx}
\usepackage{bm}

\begin{document}


\title{Distinguishing Newly Born Strange Stars from Neutron Stars with $g$-Mode Oscillations}

\author{{Wei-jie Fu$^{a}$, Hai-qing Wei$^{b}$,
and Yu-xin Liu$^{a,c,}$\thanks{Corresponding author, e-mail address: yxliu@pku.edu.cn} }\\[3mm]
\normalsize{$^a$ Department of Physics and State Key Laboratory of
Nuclear Physics and Technology,}\\
\normalsize{ Peking University, Beijing 100871, China}\\
\normalsize{$^b$ oLambda, Inc., Sunnyvale, CA 94089, USA } \\
\normalsize{$^c$ Center of Theoretical Nuclear Physics, National
Laboratory of Heavy Ion Accelerator,}\\ \normalsize{ Lanzhou 730000,
China}  }

\maketitle

%
%


\begin{abstract}
The gravity-mode ($g$-mode) eigenfrequencies of newly born strange
quark stars (SQSs) and neutron stars (NSs) are studied. It is found
that the eigenfrequencies in SQSs are much lower than those in NSs
by almost one order of magnitude, since the components of a SQS are
all extremely relativistic particles while nucleons in a NS are
non-relativistic. We therefore propose that newly born SQSs can be
distinguished from the NSs by detecting the eigenfrequencies of the
$g$-mode pulsations of supernovae cores through gravitational
radiation by LIGO-class detectors.
\end{abstract}





\newpage

\parskip=1mm

Since their discovery forty years ago, pulsars have been serving as
unprecedented astro-laboratories, especially due to their extreme
density and magnetic field. However, the basic nature of pulsar-like
stars remains murky, since, besides the conventional neutron stars
(NSs), strange quark stars (SQSs) are also potential candidates for
the
stars~\cite{Itoh1970,Freedman1978,Witten1984,Alcock1986,Weber2005,Alford2007,Jaikumar2007},
due to the quark deconfinement phase transition at high density.
More recently, the composition of pulsars, especially the existence
of SQSs, has become a hotly debated topic (see, for example,
Refs.~\cite{Alford2007,Jaikumar2007,Ozel2006,Walter2006,Lattimer2007}),
since the observation of the object PSR J0751+1807 yielded a mass
$(2.1\pm 0.2)\, \mbox{M}_\odot$ with $1\sigma$ error
bars~\cite{Nice2005}, and one of the pulsars Ter 5 I and J has a
mass larger than $1.68\, M_\odot$ to $95\%$
confidence~\cite{Ransom2005}. However, in general view, the equation
of state (EOS) of quark matter is much softer, the quark star may
not be so massive~\cite{Ozel2006}. The composition of pulsars and
the method to distinguish the SQSs from NSs becomes a fundamental
topic not only in astrophysics but also in high energy physics and
nuclear physics.

Many potentially observable differences have been proposed to
distinguish SQSs from NSs. For example, SQSs have much larger
dissipation rate of radial vibrations~\cite{Wang1984} and higher
bulk viscosity~\cite{Haensel1989}; the spin rate of SQSs can be much
closer to the Kepler limit than that of NSs~\cite{Madsen1992}; SQSs
may cool more rapidly than NSs within the first $30$
years~\cite{Schaab1997}. In this letter, we propose that the
eigenfrequencies of the gravity-mode ($g$-mode) pulsations of
supernova cores provide another sharp criteria to distinguish SQSs
from NSs after their birth in a core-collapse supernovae.

Recent simulations of core-collapse supernovae have indicated that
core $g$-mode pulsations may be excited by turbulence and downstream
accretion~\cite{Burrows2006}, due to the stalled supernova shock
undergoing a standing-accretion-shock instability and becoming
highly deformed~\cite{Foglizzo2001,Blondin2003}. Following this
discovery, Ott {\it et al}. have found that these $g$-mode
pulsations of supernova cores serve as efficient sources of
gravitational waves~\cite{Ott2006}. It is natural to expect that
more information about the interior constituents of supernova cores
carried by gravitational waves, such as whether SQSs can be formed
during the processes of supernovae, could be obtained by groundbased
detectors such as LIGO and VIRGO. In this letter we present our
calculations of the $g$-mode eigenfrequencies for newly born NSs and
SQSs in the supernova cores. Our results show that the frequencies
of newly born SQSs are lower than those of NSs by as much as one
order of magnitude. Such a large difference may provide an
appropriate means to distinguish SQSs from NSs that are newly formed
in a core-collapse supernovae. The present work is then of general
interest in nuclear/quark physics, gravitation-wave physics and
supernova physics.

Our calculations start with the differential equations in the
Newtonian formulation which governs the linear oscillations of a
nonrotating, unmagnetized, and fluidic star (for details, see for
example, Refs.~\cite{Cox1980,Reisenegger92,Lai1994}). If the
equilibrium configuration of a star is assumed to be spherically
symmetric, the pressure $p_{0}$, the density $\rho_{0}$, and the
gravitational potential $\phi_{0}$ depend only on the radial
coordinate $r$. Oscillations of the star can be described by a
vector field $\bm{\xi}(\bm{r},t)$ of displacement and the Eulerian
(or ``local") perturbations of the pressure, density, and the
gravitational potential are: $\delta p$, $\delta \rho$, and $\delta
\phi$. We adopt the Cowling approximation, i.e., neglecting
$\delta\phi$~\cite{Cowling1941}. Let the radial component of
$\bm{\xi}(\bm{r},t)$ be
$\xi_{r}Y_{lm}(\theta,\varphi)\mathrm{e}^{-\mathrm{i}\omega t}$,
following the equations of motion and continuity we have the linear
oscillation equations as
\begin{eqnarray}
\frac{d}{d r}(r^{2}\xi_{r})\! & = &
\frac{g}{c_{s}^{2}}(r^{2}\xi_{r})
+\left[\frac{l(l+1)}{\omega^{2}}-\frac{r^{2}}{c_{s}^{2}}\right]
\left(\frac{\delta p}{\rho_{0}}\right) \, , \label{Oscil1}
\\
\frac{d}{d r}\left(\frac{\delta p}{\rho_{0}}\right)\! & = &
\frac{\omega^{2}-\omega_{BV}^{2}}{r^{2}}(r^{2}\xi_{r})
+\frac{\omega_{BV}^{2}}{g}\left(\frac{\delta p}{\rho_{0}}\right) \,
, \label{Oscil2}
\end{eqnarray}
where the dependence of $\delta p$ on the spherical harmonics
$Y_{lm}(\theta,\varphi)$ and the time factor
$\mathrm{e}^{-\mathrm{i}\omega t}$ is separated analogous to
$\xi_{r}$; $\omega$ is the eigenfrequency of a mode with the index
$\alpha\equiv\{n,\,l,\,m\}$ being abbreviated; $g$ is the local
gravitational acceleration, and the Brunt-V\"{a}is\"{a}l\"{a}
frequency $\omega_{BV}$ is defined as
\begin{equation}
\omega_{BV}^{2}=g^{2}\left(\frac{1}{c_{e}^{2}}-\frac{1}{c_{s}^{2}}\right)
\, ,\label{BV}
\end{equation}
with $c_{s}$, $c_{e}$ being, respectively, the adiabatic, the
equilibrium sound speeds defined as
\begin{equation}
c_{s}^{2}\equiv\left(\frac{\partial p}{\partial
\rho}\right)_{\mathrm{adia}},\quad\quad c_{e}^{2}\equiv\left(\frac{d
p}{d \rho}\right)_{\mathrm{equi}}=\frac{d p_{0}/d r}{d \rho_{0}/d
r}\,.\label{cs}
\end{equation}
Eqs. (\ref{Oscil1}) and (\ref{Oscil2}) determine the eigenfrequency
$\omega$ for the modes, when complemented by proper boundary
conditions:
\begin{equation}
r^{2}\xi_{r}=\frac{l}{\omega^{2}}\frac{\delta p}{\rho_{0}}r\sim
r^{l+1}\qquad (r\rightarrow 0)\,,\label{bound1}
\end{equation}
at the stellar center, and
\begin{equation}
\Delta p=\delta p-\rho_{0}g\xi_{r}=0\qquad (r=R)\,,\label{bound2}
\end{equation}
for a vanishing Lagrangian perturbation of pressure at the stellar
surface $r=R$.

The equilibrium configurations of newly born NSs and SQSs are
determined by solving the Newtonian hydrostatic equations, which are
consistent with Eqs.~(\ref{Oscil1}) and (\ref{Oscil2}). The sound
speeds $c_{s}$ and $c_{e}$, thus the Brunt-V\"{a}is\"{a}l\"{a}
frequency $\omega_{BV}$, are calculated as functions of the radius
$r$. Eqs.~(\ref{Oscil1}) and (\ref{Oscil2}) are numerically
integrated from a small value of $r$ approaching to zero to the
stellar surface $r=R$. The outer boundary condition of
Eq.~(\ref{bound2}) is enforced by a shooting routine. The
eigenfrequency $\omega$ is computed iteratively after a trial value.
The order $n$ of a mode counts the number of nodes in the radial
displacement $\xi_{r}$.

To proceed, a NS is assumed to consist of nucleons, electrons,
thermal photons and neutrinos trapped in the dense matter, in the
first tens of seconds after the core bounce in a core-collapse
supernova \cite{Bethe1990}. A fluid element inside the star is
characterized by the baryon density $\rho_{\mathrm{B}}$, the entropy
per baryon $S$, and the lepton fraction
$Y_{\mathrm{L}}=Y_{\mathrm{e}}+Y_{\nu_{\mathrm{e}}}$ (or $Y_{e}$
related with $Y_{\mathrm{L}}$ through beta equilibrium conditions,
here $Y_{\mathrm{i}}=\rho_{\mathrm{i}}/\rho_{\mathrm{B}}$), all
depending on the radial coordinate $r$. In this work all the
components except the nucleons are assumed to be non-interacting,
while the interactions between nucleons are described by the
relativistic mean field (RMF) theory at finite temperature (see, for
example, Refs.~\cite{Serot86,Glendenning00}). We take the formalism
of the Lagrangian density for RMF as the same as that used in
Ref.~\cite{Fu08}. Five parameters for the theory are fixed by
fitting the properties of the symmetric nuclear matter at saturation
density: saturation nucleon number density
$\rho_{0}=0.16\:\mathrm{fm^{-3}}$, binding energy
$E/A=-16\:\mathrm{MeV}$, nucleon effective mass $m^{*}=0.75\,m_{N}$
(the bare mass $m_{N}=938\:\mathrm{MeV}$), incompressibility
$K=240\:\mathrm{MeV}$ and the symmetry energy
$E_{s}=30.5\:\mathrm{MeV}$.

Within the framework of RMF theory, we can obtain the energy density
and the pressure of the hot NS matter. In the absence of neutrino
diffusion, the adiabatic sound speed reads
\begin{equation}
c_{s}^{2}=\left(\frac{\partial p}{\partial
\rho}\right)_{S,\,Y_{\mathrm{L}}}\,.\label{cs2}
\end{equation}
It is remarkable that, in the interior of a newly born NS, the
temperature is of order $10\:\mathrm{MeV}$ and the timescale of
establishing the beta equilibrium can be smaller than
$\sim\!10^{-8}\:\mathrm{s}$~\cite{Fu08}, which is much smaller than
the period of the $g$-mode pulsations about
$10^{-3}\!\sim\!10^{-2}\:\mathrm{s}$. Therefore, in our calculation
we assume that $Y_{\mathrm{L}}$ for each fluid element is constant
and the system is in beta equilibrium during the process of
pulsations, which is consistent with the Ledoux convective criterion
in supernova simulations~\cite{Buras2006}. This situation is
different from that in cold NS where beta equilibrium can not be
obtained during the pulsations and the electron fraction
$Y_{\mathrm{e}}$ rather than $Y_{\mathrm{L}}$ remains
unchanged~\cite{Reisenegger92}. The equilibrium sound speed $c_{e}$
can also be fixed for an equilibrium configuration, where
$\rho_{\mathrm{B}}$, $S$, and $Y_{\mathrm{L}}$ ($Y_{e}$) are simply
functions of the radius $r$.

For a newly born SQS, it is usually believed to be composed of three
flavors of quarks (u, d and s), electrons, three flavor neutrinos,
thermal photons, and gluons. In our present calculation, along the
line of Ref.~\cite{Farhi1984}, we implement the MIT bag
model~\cite{Chodos1974} for its EOS. We choose
$m_{\mathrm{u}}=m_{\mathrm{d}}=0$,
$m_{\mathrm{s}}=150\,\mathrm{MeV}$, and a bag constant
$B^{1/4}=154.5\,\mathrm{MeV}$, yielding a binding energy per baryon
of $928\,\mathrm{MeV}$ for the SQS matter in equilibrium ($p=0$),
which is comparable to $931\,\mathrm{MeV}$ per nucleon for the
${^{56}}\mathrm{Fe}$ nucleus \cite{Farhi1984}. Like in the NS case,
all the thermodynamical properties and the two sound speeds can be
determined as the equilibrium configuration of a SQS is known.

In our calculation we employ the equilibrium configuration of newly
born NSs resulting from the 2D hydrodynamic simulations of
core-collapse supernovae by the Arizona Group~\cite{Dessart2006}. We
calculate then the properties of the $g$-mode oscillations of newly
born NSs at the time $t=100\,\mathrm{ms}$, $200\,\mathrm{ms}$ and
$300\,\mathrm{ms}$ after the core bounce, along the line taken in
Ref.~\cite{Dessart2006}. The time range is limited to within
$300\,\mathrm{ms}$ for simplicity since this is guaranteed by the
fact that during the first second after the core bounce, the
eigenfrequencies vary only a little, for example, the frequencies of
$g$-modes vary within $727\,\mathrm{Hz}\sim 819\,\mathrm{Hz}$ when
the time ranges from $0.3\,\mathrm{s}$ to $1\,\mathrm{s}$ in a newly
born NS model~\cite{Ferrari2003}.

Eqs.~(\ref{Oscil1}) and (\ref{Oscil2}) are integrated from the
center to a radius of about $20\,\mathrm{km}$ where a convective
instable region appears. We choose the mass inside the radius of
$20\,\mathrm{km}$ to be $0.8\,\mathrm{M}_{\odot}$,
$0.95\,\mathrm{M}_{\odot}$ and $1.05\,\mathrm{M}_{\odot}$ at
$t\!=\!100\,\mathrm{ms}$, $200\,\mathrm{ms}$ and $300\,\mathrm{ms}$
after the core bounce, respectively, according to the supernovae
simulations~\cite{Dessart2006}. The increase of the mass with time
is due to accretions of mantle materials of the progenitor star onto
the newly born neutron star.

Table~\ref{FrqCaltd} lists the calculated $g$-mode oscillation
eigenfrequencies of an above-modeled newly born NS with order
$n=1,\, 2, \, 3$. Here the harmonic index is fixed at $l=2$, which
represents the quadrupole motion of mass and is directly related to
the emission of gravitational waves (see, for example,
Refs.~\cite{Misner70,Owen2005,Lai2006}). Our results of
$717.6\,\mathrm{Hz}$, $774.6\,\mathrm{Hz}$, and $780.3\,\mathrm{Hz}$
for $n=1$ at $t=100\,\mathrm{ms}$, $200\,\mathrm{ms}$, and
$300\,\mathrm{ms}$ after the core bounce, respectively, are
consistent with the Newtonian supernovae simulation results of about
$675\,\mathrm{Hz}$ in Ref.~\cite{Burrows2006} and about
$950\,\mathrm{Hz}$ in Ref.~\cite{Ott2006}. They are also close to
the general relativistic results ranging from $\sim
620\,\mathrm{Hz}$ to $\sim 820\,\mathrm{Hz}$ in the first second
after the core bounce~\cite{Ferrari2003}. The quantitative
differences result from different equations of state and NS models
used, general relativistic effects and so on.

\begin{table}[htb]
\begin{center}
\caption{Calculated eigenfrequencies (in unit of Hz) of the $g$-mode
oscillations of the newly born NSs and SQSs at several time $t$ (in
unit of millisecond) after the core bounce.} \label{FrqCaltd}
\begin{tabular}{|c|c|c|c|c|c|c|}
\hline \hline                   
Radial order &\multicolumn{3}{|c|}{Neutron Star} &
\multicolumn{3}{|c|}{Strange Quark Star}  \\ \cline{2-7}
of $g$-mode & $t\!=\!100$&$t\!=\!200$&$t\!=\!300$&$t\!=\!100$&$t\!=\!200$&$t\!=\!300$\\
\hline $n=1$ & 717.6 & 774.6 & 780.3 & 82.3 & 78.0 & 63.1 \\
\hline $n=2$ & 443.5& 467.3 & 464.2 &  52.6 & 45.5 & 40.0   \\
\hline $n=3$ & 323.8& 339.0 & 337.5 &  35.3 & 30.8 & 27.8   \\
\hline \hline
\end{tabular}
\end{center}
\end{table}

For comparison, we proceed to calculate the $g$-mode oscillation
eigenfrequencies of newly born SQSs. Despite the lack of commonly
accepted simulations on supernovae with strange quark cores in
literatures, following the results of the trial
simulations~\cite{Lin2006,Yasutake2007}, general theoretical
investigations\cite{Jaikumar2006,Haskell2007} and analysis of some
supernovae observations~\cite{Chen2007,Leahy2008} we assume that a
strange quark core can be formed during the collapse of
presupernova.

To carry out the calculation for SQSs, we assume that the radial
variation behaviors of $S$ and $Y_{\mathrm{L}}$ for newly born SQSs
are the same as for NSs. But a newly born SQS differs from a NS in
its smaller size, having a radius of about $10\,\mathrm{km}$ where
the pressure vanishes. With the same masses as a NS at the three
time instants after the core bounce, we calculate the $l\!=\!2$
eigenfrequencies of a newly born SQS at first. The obtained results
are also listed in Table~\ref{FrqCaltd}. It is apparent that the
$g$-mode oscillation frequencies of $82.3\,\mathrm{Hz}$,
$78.0\,\mathrm{Hz}$, and $63.1\,\mathrm{Hz}$ of a SQS are definitely
lower than those of a NS by an order of magnitude.
Table~\ref{FrqCaltd} also shows the frequencies of modes with radial
order higher than one, although such higher-order $g$-mode
oscillations are less likely to be significant sources of
gravitational waves, due to the much longer gravitational wave
damping time and thus much weaker coupling to gravitational
waves~\cite{Ferrari2003}. We have also performed calculations in the
case of that the entropy per baryon $S$ is increased to $1.5S$ at
each fluid element inside the SQS, by considering the release of
thermal energy during the conversion from nuclear matter to strange
quark matter~\cite{Anand1997}. The eigenfrequencies of the $n=1$
$g$-mode oscillations become $100.7\,\mathrm{Hz}$,
$105.9\,\mathrm{Hz}$, and $96.1\,\mathrm{Hz}$ at the same three time
instants, respectively, which are still significantly lower than
those of a newly born NS. The dependence of the $g$-mode frequencies
on the mass of a SQS has also been investigated. Our finding is that
the eigenfrequencies of a newly born SQS increase a little for a
larger mass. For example, a SQS core with mass
$1.4\,\mathrm{M}_{\odot}$ holds the $n=1$ order $g$-mode frequencies
$100.2\,\mathrm{Hz}$, $91.4\,\mathrm{Hz}$, and $73.0\,\mathrm{Hz}$
at the three instants, respectively. And the frequencies change to
$108.8\,\mathrm{Hz}$, $100.9\,\mathrm{Hz}$, and $84.5\,\mathrm{Hz}$
accordingly as the mass of the SQS is increased to
$1.68\,\mathrm{M}_{\odot}$. They are also much lower than those of
newly born NSs.

\begin{figure}
\centering
\includegraphics[scale=1.0,angle=0]{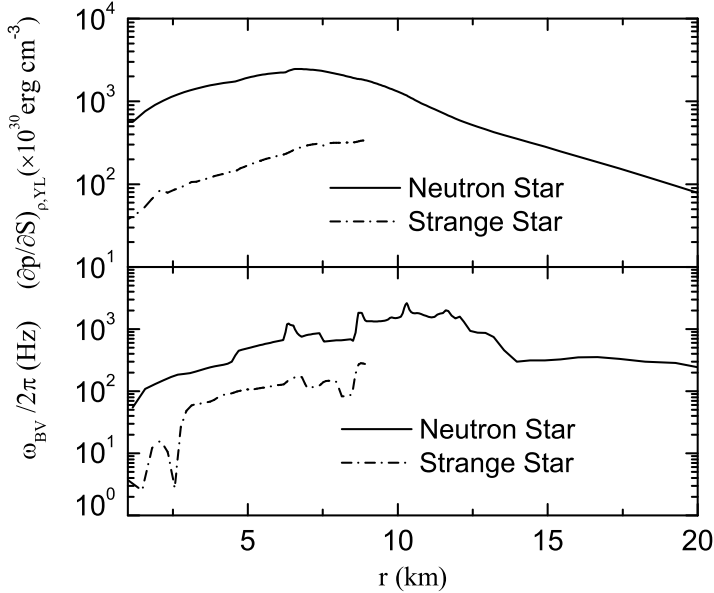}
\vspace*{-8mm}\caption{\label{BVpic} Calculated $(\partial
p/\partial S)_{\rho,Y_{L}}$ (top) and Brunt-V\"{a}is\"{a}l\"{a}
frequency $\omega_{BV}$ (bottom) as functions of the radial
coordinate at $t\!=\! 200\mathrm{ms}$ after the core bounce for
newly born NSs and SQSs. A sharp surface appears at about
$r=9\,\mathrm{km}$ for the SQS.}
\end{figure}

To explore the reason for the large difference in the $g$-mode
oscillation eigenfrequencies between newly born NSs and SQSs, we
reexpress the Brunt-V\"{a}is\"{a}l\"{a} frequency in Eq.~(\ref{BV})
as
\begin{equation}
\omega_{BV}^{2}=-\frac{g}{\rho} \left[\left(\frac{\partial
\rho}{\partial Y_{L}}\right)_{p,S}\frac{d Y_{L}}{d
r}+\left(\frac{\partial \rho}{\partial S}\right)_{p,Y_{L}}\frac{d
S}{d r}\right] \, ,\label{BV2}
\end{equation}
where we further have
\begin{eqnarray}
\left(\frac{\partial \rho}{\partial
Y_{L}}\right)_{p,S}=-\frac{1}{c_{s}^{2}}\left(\frac{\partial
p}{\partial Y_{L}}\right)_{\rho,S}\,,\label{rhoS}\\
\left(\frac{\partial \rho}{\partial
S}\right)_{p,Y_{L}}=-\frac{1}{c_{s}^{2}}\left(\frac{\partial
p}{\partial S}\right)_{\rho,Y_{L}}\,.\label{rhoYL}
\end{eqnarray}
Comparing the calculations for SQSs with those for NSs, we find that
the values of the $(\partial p/\partial S)_{\rho,Y_{L}}$ in
Eq.~(\ref{rhoYL}) for SQSs are much smaller than those for NSs,
which result in the values of the Brunt-V\"{a}is\"{a}l\"{a}
frequency $\omega_{BV}$ for SQSs much smaller than those for NSs as
well. Quantitative comparisons of $(\partial p/\partial
S)_{\rho,Y_{L}}$ and $\omega_{BV}$ between SQSs and NSs are
displayed in Fig.~\ref{BVpic} for an illustrative time
$t=200\,\mathrm{ms}$. Those calculated results can be understood
easily as follows: a SQS is composed of particles which are all
extremely relativistic. In the ideal case of neglecting the mass of
the strange quark, i.e. $m_{s}=0$, the EOS of a SQS has a simple
form as $p=\rho/3-4B/3$ in the MIT model. Substituting this EOS into
Eqs.~(\ref{rhoS}) and~(\ref{rhoYL}) we find both $(\partial
p/\partial S)_{\rho,Y_{L}}$ and $(\partial p/\partial
Y_{L})_{\rho,S}$ are vanishing and thus the $g$-mode frequency of
the SQS in this extreme situation is zero. This simple analysis is
confirmed in our numerical calculations when setting $m_{s}=0$.
Therefore, the nonvanishing eigenfrequencies of $g$-mode pulsations
for a realistic SQS as given in this work are due to the nonzero
value of the strange quark mass. It should be emphasized that this
result does not depend on the details of the model for the strange
quark matter but only on the relativistic properties of the
components of SQSs. Once the matter is relativistic and the EOS can
be approximately parameterized as $p=\rho/3+const.$, the $g$-mode
frequencies are highly suppressed. While in a NS, beside the
relativistic leptons, there are non-relativistic nucleons (or,
simply, with quite large masses), which are responsible for the much
higher $g$-mode eigenfrequencies.

In summary, we have calculated the eigenfrequencies of $g$-mode
pulsations in newly born neutron stars and strange quark stars in
this letter. Special emphasis is given to the $g$-mode oscillations
with harmonic index $l=2$ and order $n=1$ due to the potential
detectability of their gravitational radiation. Because their
components are all relativistic particles, the $g$-mode oscillation
frequencies of newly born strange quark stars are around or below
$100\,\mathrm{Hz}$ for $l\!=\!2$ and $n\!=\!1$. They are definitely
much lower than those of newly born neutron stars ($\sim 650$ -
$950\, \mathrm{Hz}$, combining our results and those in literatures)
by almost one order of magnitude. Since gravitational waves and the
difference of the $g$-mode oscillation frequencies may be detected
by LIGO-class detectors, our work illustrates an important signature
to distinguish newly born strange quark stars from the neutron
stars.

\bigskip

This work was supported by the National Natural Science Foundation
of China under contract Nos. 10425521 and 10675007, the Major State
Basic Research Development Program under contract No. G2007CB815000,
the Key Grant Project of Chinese Ministry of Education under contact
No. 305001.


\end{document}